      [1999/12/01 v1.4c Il Nuovo Cimento]
\documentclass{cimento}

\usepackage{graphicx}  
\title{Top BSM at D0}
\author{Daniel Wicke\from{ins:W}
}
\instlist{\inst{ins:W} Bergische Universit{\"a}t Wuppertal \-
  Gau\ss{}str. 20, 42097 Wuppertal\\
  for the D0 collaboration
}
\PACSes{%
\PACSit{14.65.Ha}{Top quarks}
\PACSit{14.70.Pw}{Other gauge bosons}
\PACSit{14.80.-j}{Other particles (including hypothetical)}
}
\newlength{\ziffer}

\newcommand{\TeV}{\,\mbox{Te\kern-0.2exV}}
\newcommand{\GeV}{\,\mbox{Ge\kern-0.2exV}}
\newcommand{\mGeV}{\,\mathrm{Ge\kern-0.2exV}}
\newcommand{\MeV}{\,\mbox{Me\kern-0.2exV}}
\newcommand{\keV}{\,\mbox{ke\kern-0.2exV}}
\newcommand{\eV}{\,\mbox{e\kern-0.2exV}}

\newcommand{\ifb}{\,\mbox{fb}^{-1}}
\newcommand{\pb}{\,\mbox{pb}}

\newcommand{\bea}{\pagebreak[3]\begin{samepage}\begin{eqnarray}}
\newcommand{\eea}{\end{eqnarray}\end{samepage}\pagebreak[3]}
\newcommand{\beq}{\begin{equation}}
\newcommand{\eeq}{\end{equation}}

\newcommand{\met}{\slash\!\!\!\!{E_T}}

\newcommand{\abb}{Fig.~\ref}
\newcommand{\fig}{\abb}

 \newlength{\howlong}

\begin{document}

\maketitle

\begin{abstract}
The D0 experiment has searched for phenomena beyond the standard model
in top quark events. The methods and results of four analyses covering various possible
deviations from the standard model behaviour are discussed. With data
sets covering up to $2.1\ifb$  no deviation from the
standard model expectations could be found.
\end{abstract}

\section{Introduction}
Since the top was discovered by CDF and D0 at the Tevatron in 1995~\cite{Abe:1995hr,Abachi:1995iq} the
number of top events available for experimental studies has been
increased by more than an order of magnitude. Tevatron now delivered
a luminosity of more than $4\ifb$ up to half of which has been
analysed for top quark analyses in D0. These data are, amongst other studies, investigated to
verify whether the events selected as top quarks  actually behave as expected 
by the Standard Model (SM). 

The questions one may ask to challenge the SM-likeness of
the top quark naturally fall into 4 categories. First, one may
ask whether the events that are considered to be top quarks  actually are
all top quarks or whether some additional unknown new particle is hiding in
the selected data. 
Second one may ask whether the top quark decay looks as predicted by
the SM, maybe a contribution from a new particle can be detected. Third one may ask whether the
quantum numbers of the top are the expected ones or whether exotic
scenarios play a role.
And fourth one may ask if non-standard production mechanisms add
to the SM diagrams.

The analyses considered address these questions within the selections used to detect one of
the SM decay channels of the top quark. For top pair production these
are defined by the decay of two $W$ bosons as dilepton, $\ell+$jets
and all hadronic channel.
This writeup covers one D0 analysis for each of the above
questions using data selected in one or more of the top pair decay channels. 
Additional analyses of top events with 
interpretations beyond the standard model are discussed elsewhere in
these proceedings~\cite{shabnam,lisa}.

\section{Search for Stop Admixture}
A particle that may hide in the samples usually considered to
be top quarks are its supersymmetric partners, the stop quarks, $\tilde{t}_1$ and
$\tilde{t}_2$. The
stop decay modes to neutralino and top quark, $\tilde\chi_1^0 t$, or through chargino and
$b$ quark, $\tilde\chi_1^\pm b$, both yield neutralino, $b$ quark and $W$ boson,
$\tilde\chi_1^0 b W$. The  neutralino is the lightest supersymmetric
particle in many models and is stable if $R$-parity in conserved. 
Then it escapes the detector and the experimental signature of  stop pair
production differs from that of top pair production only by the 
additional contribution to the  missing transverse energy from the
neutralino.
\subsection{Data Selection, Signal and Background Description}\label{sect:stop.selection}
D0 has searched for a contribution of such stop pair production in
the semileptonic channel in data with \mbox{$\sim\!0.9\ifb$}\cite{d0note5438conf}. Semileptonic
events were selected following the corresponding $t\bar t$
cross-section analysis by looking for isolated leptons ($e$ and $\mu$),
missing transverse energy and four or more jets. At least one of the jets
was required to be identified as $b$-jet using D0's neural network algorithm. 
To describe the SM expectation a mixture of data and simulation (MC)
is employed.  The description of top pair production (and of further
minor backgrounds) is taken fully from MC normalised to the
corresponding theoretical cross-sections. For $W+$jets the kinematics
is taken from MC, but the normalisation is taken from data. 
Multijet background is fully estimated from data.  
As signal we consider the lighter of the two stop quarks, $\tilde{t}_1$.
Production of $\tilde{t}_1 \bar{\tilde{t}}_1$ is simulated for various
combinations of stop and chargino masses, $m_{\tilde{t}_1}, m_{\tilde\chi^\pm_1}$. For the sake of this
analysis the stop mass was chosen to be less or equal to the top mass. The
neutralino mass $m_{\tilde\chi^0_1}=50\GeV$ was chosen to be close
to the experimental limit.

\subsection{Signal Background Separation} 
\begin{figure}[b]
  \centering
\includegraphics[width=0.49\linewidth]{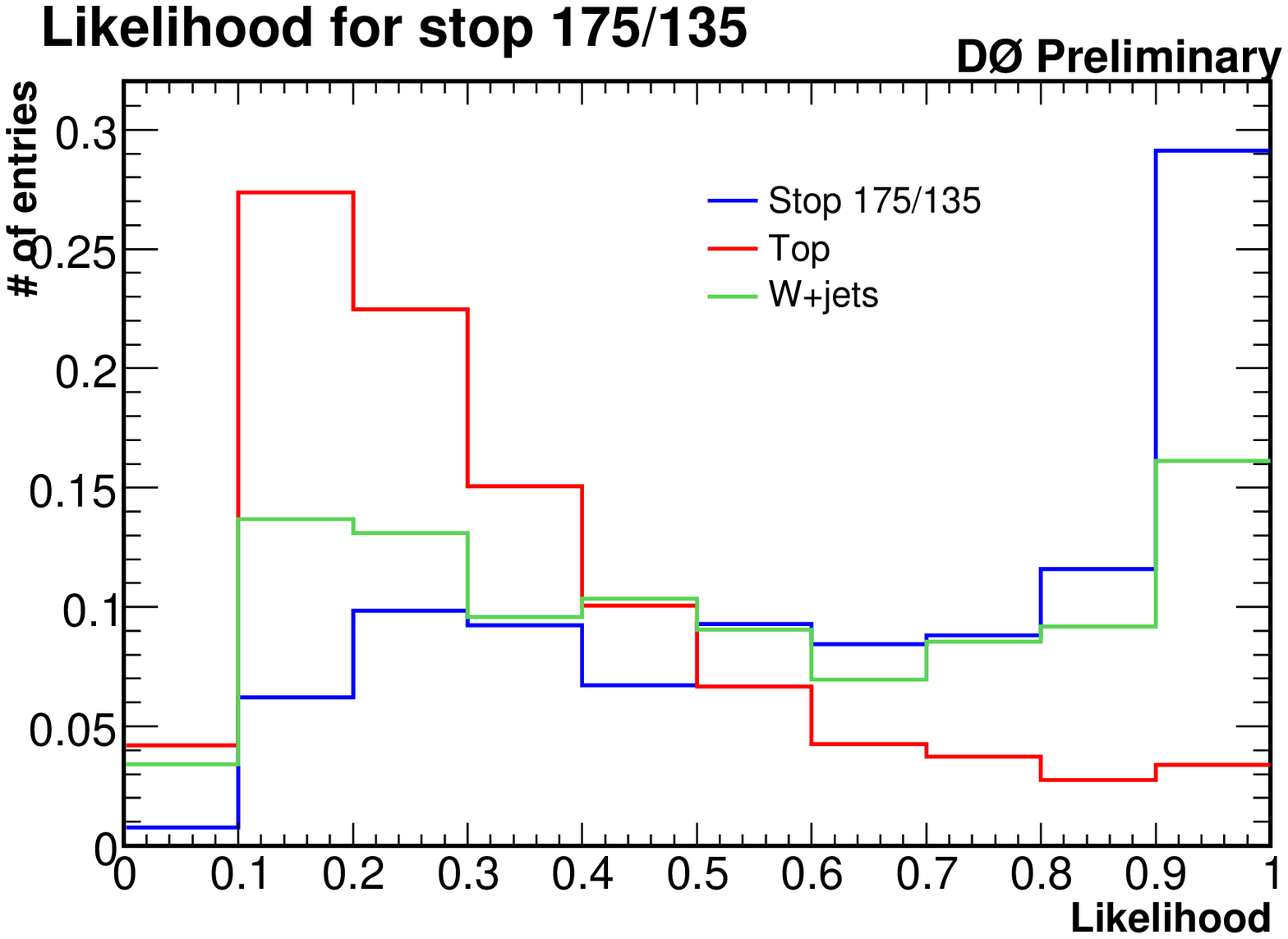}
\includegraphics[width=0.49\linewidth]{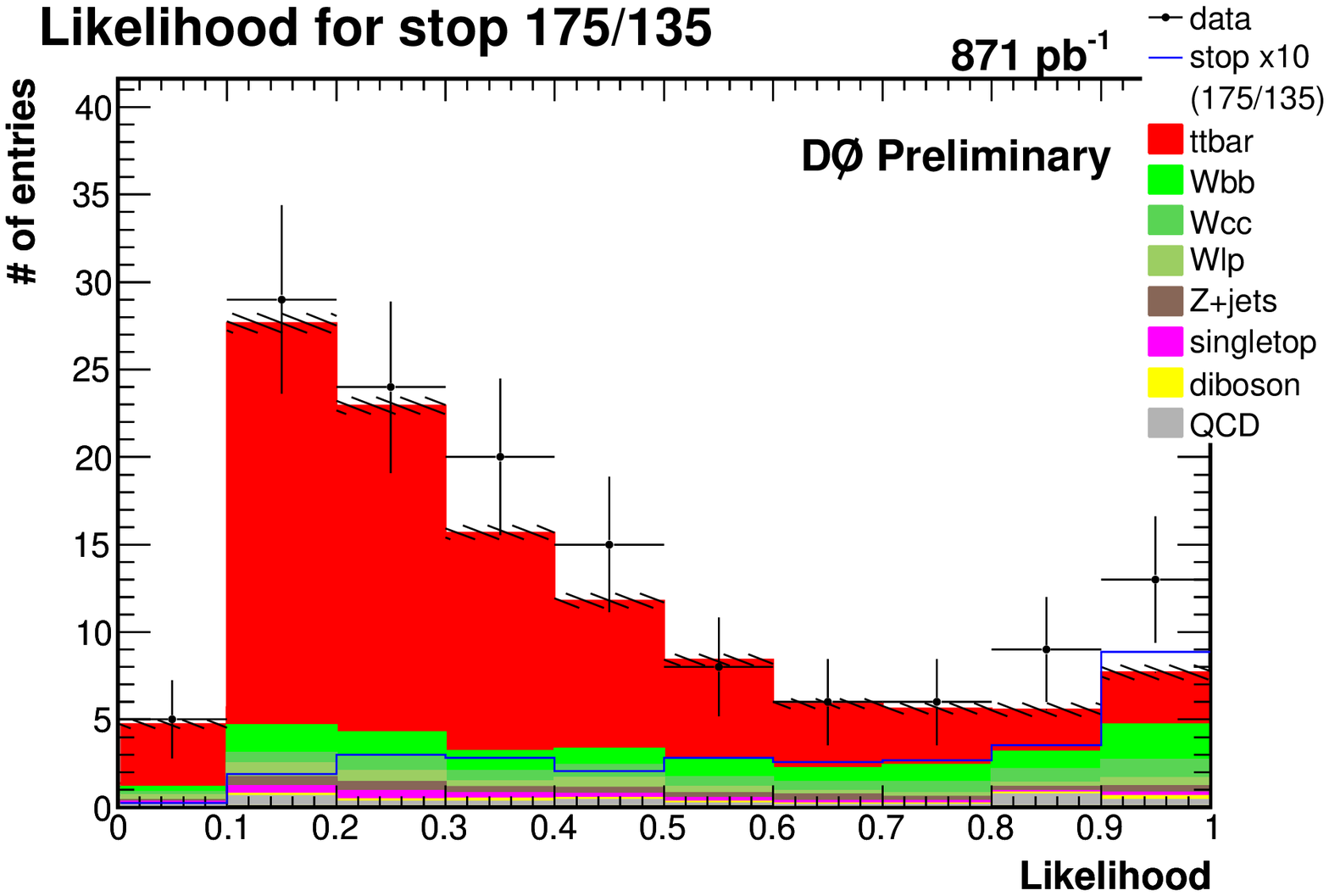}%
  \caption{\label{fig:stop.likelihood}%
Likelihood distribution $m_{\tilde{t}_1}=175\GeV$, $m_{\tilde\chi_1^\pm}=135\GeV$.  
Left: Expected shape for stop signal, SM top pairs and $W+$jets events.
Right: Expected distribution for SM compared to data. The stop
contribution expected in the MSSM is enhanced by a factor of 10.
}
\end{figure}
To detect a possible contribution of
stop pairs the differences between stop pair events and SM top pair production  
kinematic variables are combined into a likelihood, 
${\cal  L}=P_\mathrm{stop}/\left(P_\mathrm{stop}+P_\mathrm{SM}\right)$.
The kinematic variables considered include the transverse momentum of
the (leading) $b$ jet, distances between leading $b$ jet and lepton or
leading other jet. Additional variables were reconstructed by applying
a constraint fit.  In this fit reconstructed physics objects (lepton,
missing transverse energy and jets) are assigned to
the decay products of an assumed semileptonic top pair event and 
the measured quantities are allowed to vary within their experimental
resolution to fulfil additional constraints. It was required that the $W$ mass is consistent with the invariant mass
of the jets assigned to the two light quarks as well as with the
invariant mass of the lepton with the neutrino.
The masses of the reconstructed top pairs were constrained to be equal.
Of the possible jet parton assignments only the one with the best
$\chi^2$ was used. 
From the constraint fits observables the angle between the $b$-quarks and the
beam axis in the $b\bar b$ rest frame, the $b\bar b$ 
invariant mass, the distances between the $b$'s and
the same-side or opposite-side $W$ bosons and the reconstructed top
mass are considered. 
The likelihood was derived  for each $m_{\tilde{t}_1},
m_{\tilde\chi^\pm_1}$ combination separately and the selection of variables used
has been optimised each time. Figure~\ref{fig:stop.likelihood}
shows the separation power of the likelihood for the case of
$m_{\tilde{t}_1}=175\GeV$, $m_{\tilde\chi^\pm_1}=135\GeV$ and the comparison to the observed data.

\subsection{Limits and Cross Checks}\label{sect:stop.limit}
\begin{figure}[t]
  \centering
\includegraphics[width=\linewidth]{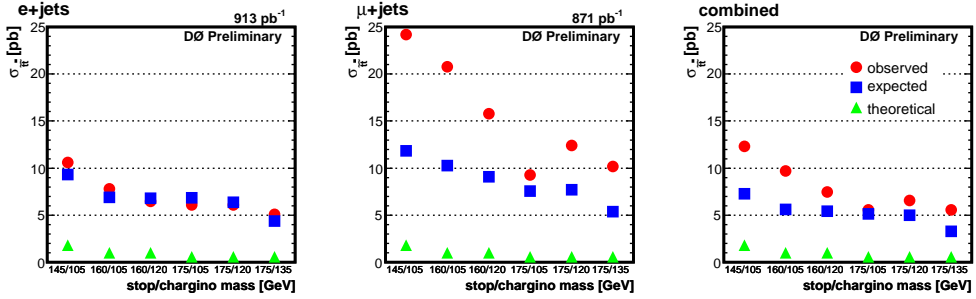}
  \caption{\label{fig:stop.limits}Expected and observed limits on the stop
    pair production cross-section compared to the expectation in the
    MSSM for $e+$jets (left), $\mu+$jets (middle) and combined
     data (right).}
\end{figure}

\begin{figure}[b]
  \centering
\includegraphics[width=\linewidth]{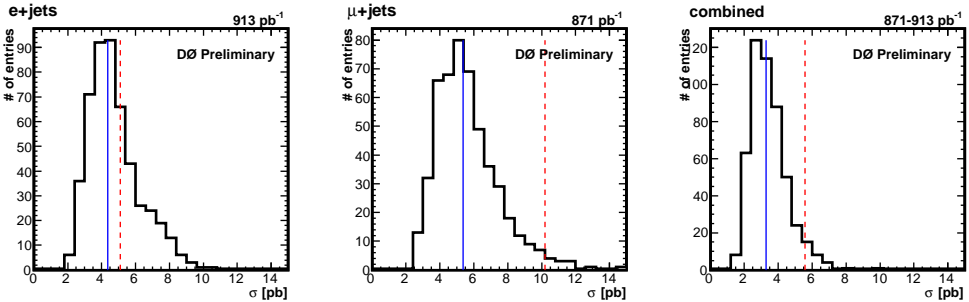}
  \caption{\label{fig:stop.ensemble}Distribution of results for  $m_{\tilde{t}}=175\GeV$,
$m_{\tilde\chi^\pm}=135\GeV$ obtains in 500~sets of SM pseudo data. The
expected and observed results are shown as full and dashed lines, respectively.}  
\end{figure}
To determine limits on the possible contribution of stop pair
production in the selected channel  Bayesian statistics is employed
using a non-zero flat prior (for positive values) of the stop pair cross-section.
A Poisson distribution is assumed for the number of events observed in
each bin of the likelihood. The prior for the combined signal
acceptance and background yields is a multivariate Gaussian with
uncertainties and correlations described by a covariance matrix. 
The systematic uncertainty is dominated by the uncertainties on the theoretical
cross-section of top pair production, on the selection efficiencies 
and the luminosity determination. 
Figure~\ref{fig:stop.limits} shows the expected and observed limits
on the stop pair production cross-section compared to the MSSM
prediction for various values of $m_{\tilde{t}_1}$ and $m_{\tilde\chi^\pm_1}$. 
While the $e+$jets channel shows agreement between observed and
expected limits, the observed limits in the $\mu+$jets channel are
weaker than the expectations. In all cases the theoretically expected
stop signal cross-section in the MSSM is much smaller than the
experimental limits. 
Still, the deviation of the observed from the expected limit in the
$\mu+$jets data may be a sign of new physics.
To estimated the consistency of the observation with the SM,  500~sets
of SM pseudo data were generated 
and the analysis was repeated on each of
these pseudo datasets. The distribution of limit results is shown in
\fig{fig:stop.ensemble} for the example of $m_{\tilde{t}_1}=175\GeV$,
$m_{\tilde\chi^\pm_1}=135\GeV$.  The distributions exhibit a strong asymmetry
towards higher limits and some percent of the pseudo data yield a
limit even higher than the one observed in real data. 
D0 thus concludes that the observed results are consistent with a
statistical fluctuation within the SM.

\section{Search for Decay to Charged Higgs}

New particles in the final state of top pair events may alter the
branching fractions to the various decay channels. Because measured top pair
production cross-sections, $\sigma_{t\bar t}$, are calculated assuming the SM branching
fraction, this would be visible by comparing $\sigma_{t\bar t}$ measured in
various channels.

In a first analysis D0 considers the case of a charged Higgs replacing
the $W$ in the top decay~\cite{D0Note5466}. The charged Higgs is assumed to decay
hadronically. Within the MSSM with its two Higgs doublets this case is relevant for low values of
$\tan \beta$. General multi-Higgs-doublet models allow such
leptophobic charged Higgs' over a large range of parameters. 
The observable employed is the cross-section ratio $R_\sigma=\sigma_{t\bar
  t}^{\ell+\mathrm{jets}}/\sigma_{t\bar t}^{\mathrm{Dilepton}}$.

\subsection{Determination of the Cross-Section Ratio}
The determination of cross-section ratio,  
$R_\sigma$, 
is
based on individual cross-section measurements in  the $\ell+$jets
and the dilepton channel, that are based on luminosity of  $\sim\!0.9\ifb$ and $\sim\!1.0\ifb$, respectively.
\begin{figure}[b]
  \centering
   \includegraphics[width=0.37\linewidth]{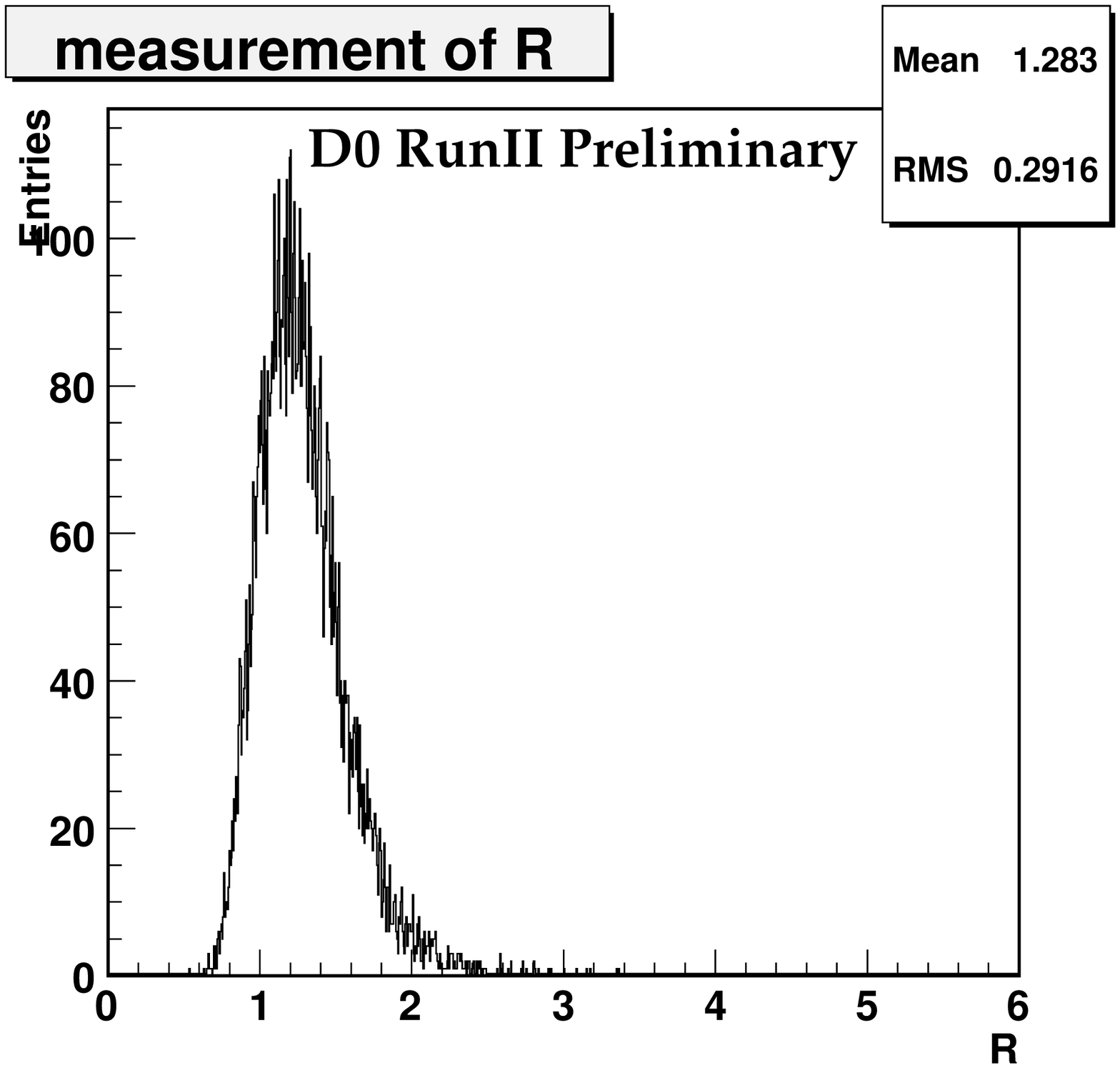} 
   \includegraphics[width=0.37\linewidth]{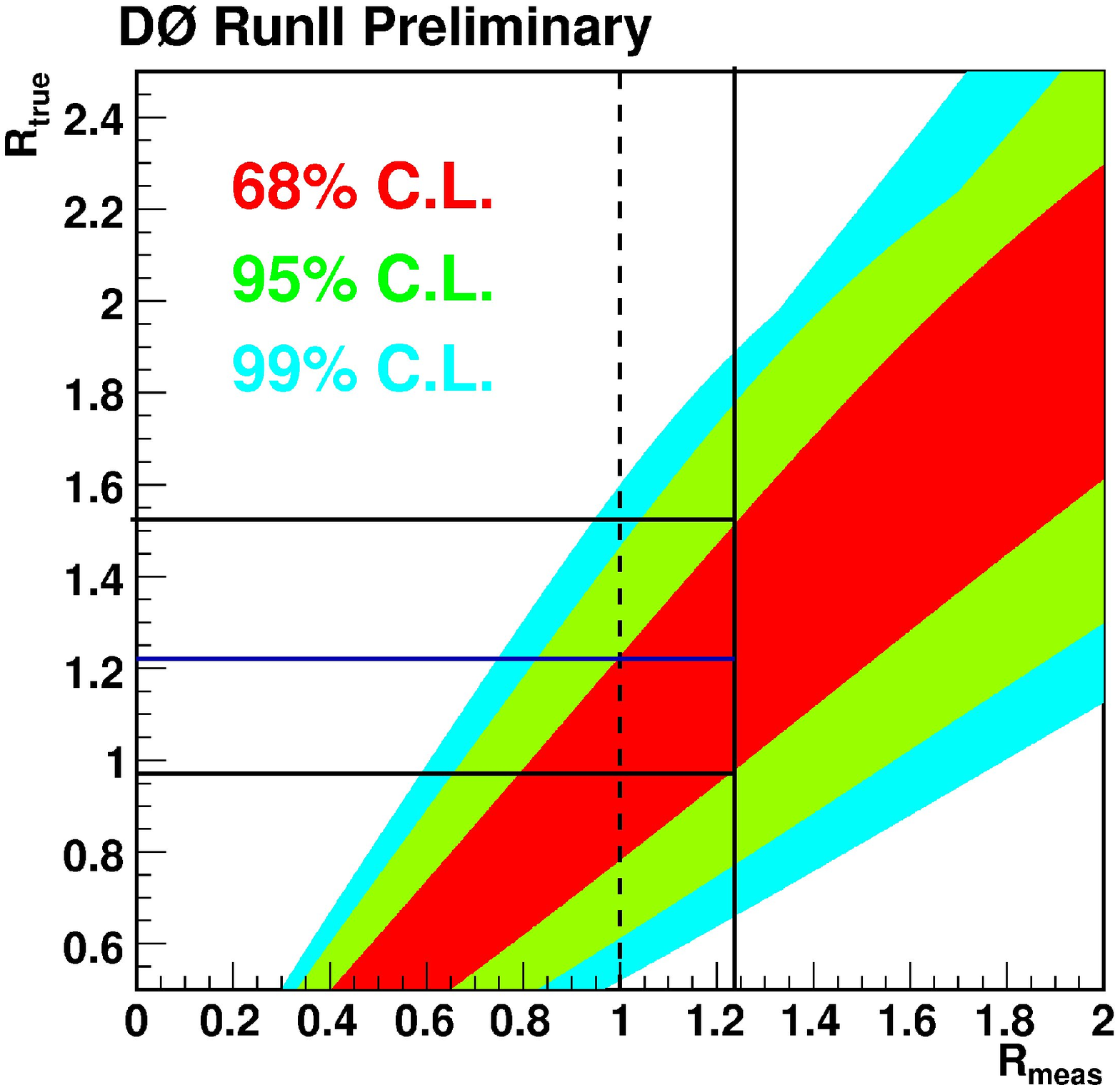}
  \caption{\label{fig:r-xsec}Left: Distribution of observed cross-section
  ratios for ensemble test with a true $R=1.2$. Right: Confidence
  intervals in the measured vs. nominal $R$ plane. }
\end{figure}
For the evaluation of the systematic uncertainties of the cross-section
ratio correlations between the channels need to be taken into
account. One class of uncertainties is considered to
be fully correlated between the channels. This class includes the 
uncertainty of efficiency for identifying leptons and the primary
vertex,  uncertainties related to jet reconstruction and of normalising diboson background simulation.
The luminosity uncertainty is cancelled in the ratio.
Remaining uncertainties of the individual measurements were treated
as being uncorrelated.
The systematic uncertainties were then computed by combined ensemble
testing. For both channels event numbers were repeatedly drawn
according to Poisson statistics. The expectation parameters were
varied according to the size of the various systematics either
correlated or uncorrelated, depending on the source of uncertainty.
This yields a distribution of results as shown in \fig{fig:r-xsec} (left).
As the outcome of the procedure depends on the assumed nominal
$R_\sigma$, it has to be repeated for many nominal values of
$R_\sigma$, which was done by modifying $\sigma_{t\bar t}^{\ell+\mathrm{jets}}$.
After smoothing the obtained distributions with functional form, the
Feldman-Cousins method was applied:
$R_\sigma = 1.21 \pm 0.27$ was obtained, c.f.~\fig{fig:r-xsec} (right). 
The result is consistent with the SM
expectation of $1.0$.

\subsection{Top Branching Ratio to Charged Higgs}
\begin{figure}[t]
  \centering
\unitlength0.01\linewidth
  \begin{picture}(0,0)
    \put(33,0){\small $B(t\rightarrow H^\pm b)$}
  \end{picture}%
\raisebox{0.335\linewidth}[0pt][0pt]{\includegraphics[height=0.49\linewidth,angle=-90,clip,trim=0mm 0mm 17mm 0mm]{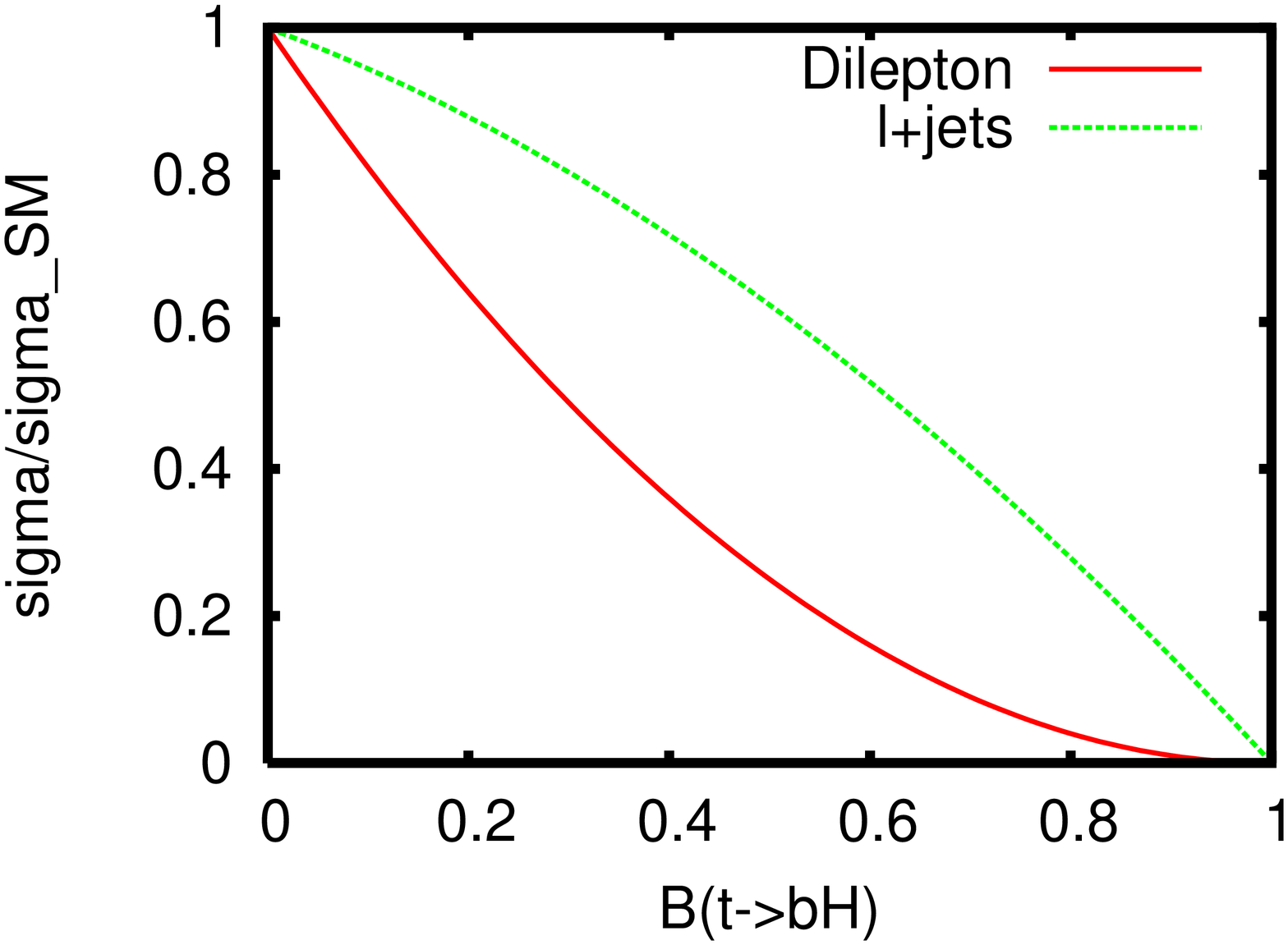}}
\includegraphics[width=0.36\linewidth]{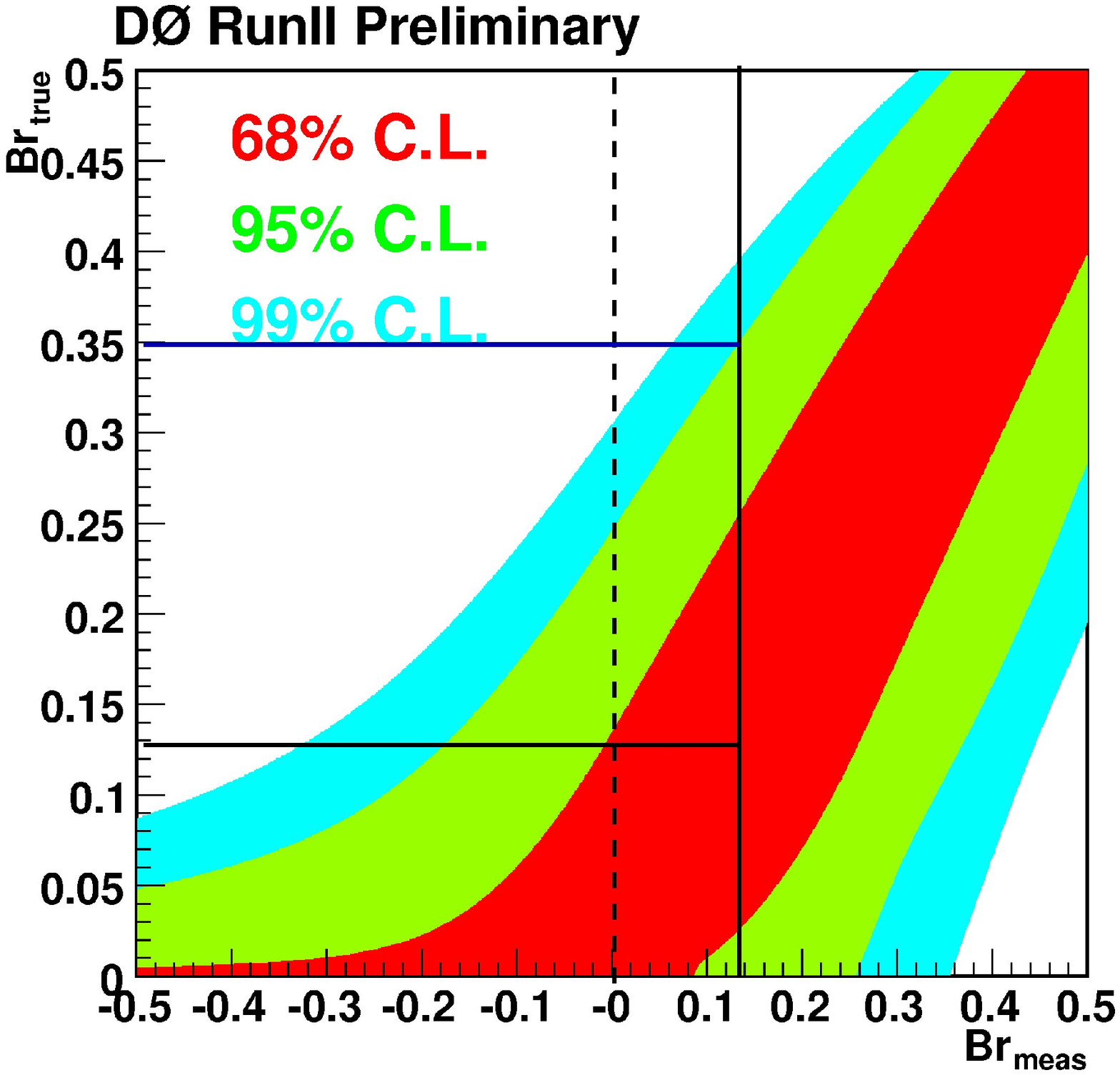}
  \caption{\label{fig:r-br} 
Left: Expected dilepton and $\ell+$jets reduction as function of top
branching fraction to leptophobic charged Higgs.
Right: Confidence  intervals in the measured vs. nominal $B$ plane.}
\end{figure}

Assuming a leptophobic charged Higgs,
i.e. $B(H^\pm\rightarrow cs)=100\%$, dileptonic events can only appear
if both top quarks decay through a $W$ boson. In $\ell+$jets events at
least one top decay must occur through a  $W$ boson.
The higher the contribution of top decays through charged Higgs the
less likely is the decay through a $W$ boson. Thus both channels will
have a reduced contribution with respect to the SM. However, the reduction is stronger in the
dilepton channel, see \fig{fig:r-br} (left). The cross-section ratio
dependence can be written as
\beq
\displaystyle R_\sigma=1+\frac{x}{(1-x) B(W\rightarrow qq)} 
\quad \mbox{with} \quad x=B(t\rightarrow H^\pm b) \mbox{.}
\eeq
As an extention to this formula the actual analysis accounts for
leakage between the channels. 
With this the above result on $R_\sigma$ can be converted
to a limit on the charged Higgs contribution in top decays.
As before ensembles of results for various nominal values of
$B(t\rightarrow H^\pm b)$ were generated with appropriate treatment of
systematic uncertainties. Then the Feldman-Cousins approach is used to
determine the final result accounting for the physical constraints of
$0\le B(t\rightarrow H^\pm b) \le1$. D0 finds  $B(t\rightarrow H^\pm
b)<0.35$ at 95\% confidence level. Because the measured $R_\sigma$ is
slightly larger than 1 the expected limit of $B(t\rightarrow H^\pm
b)<0.25$ is not reached.

\section{Search for Exotic Top Charge}
The top quark's electrical properties are fixed by its charge.
However, in reconstructing top quarks the charges of the objects
usually aren't checked. Thus an exotic charge
value of $\left|q_t\right|=4e/3$ isn't excluded by standard analyses.
To distinguish between the SM and the exotic top charge it is
necessary to reconstruct the charges of the top quark decay products,
the $W$ boson and the $b$ quark. The $W$ boson charge can be taken from the charge
of the reconstructed lepton, but finding the charge of the $b$ quark
is more difficult.

\subsection{Data Selection and Background Description}
D0 has performed an analysis of $\ell+$jets events with at least to
$b$-tagged jets in $370\ifb$ using a jet charge technique to determine the charge of the $b$ jets~\cite{Abazov:2006vd}.
Semileptonic events are selected following the cross-section analysis by requiring exactly one isolated
lepton, tranverse missing energy and 4 or more jets. At least 2 of the
jets must be identified as $b$ jets using a secondary vertex tagging
algorithm.

\subsection{Jet Charge}
The charge of a jet can be defined as the sum of the charges of all
tracks inside the cone of that jet. In this analysis the sum has been
weighted with the tracks transverse momentum:
\beq
 Q_\mathrm{jet}:=\frac{\sum{q_i\cdot p^{0.6}_{Ti}}}{\sum{p^{0.6}_{Ti}}}
\eeq
Because particles may easily escape the jet cone such a jet charge 
fluctuates strongly from event to event, only statistical
statements can be made.
It is crucial to determine the expected distribution of
$Q_\mathrm{jet}$ in the case of $b$ or $\bar b$ quark and, because a
significant fraction of charm quarks gets flagged by the secondary
vertex tagger, also for the $c$ and $\bar c$ quarks. These expected
distributions, c.f. \fig{fig:charge} (left), are derived from dijet
data using a tag and probe method. 

\begin{figure}[b]
  \centering
  \includegraphics[width=0.32\linewidth]{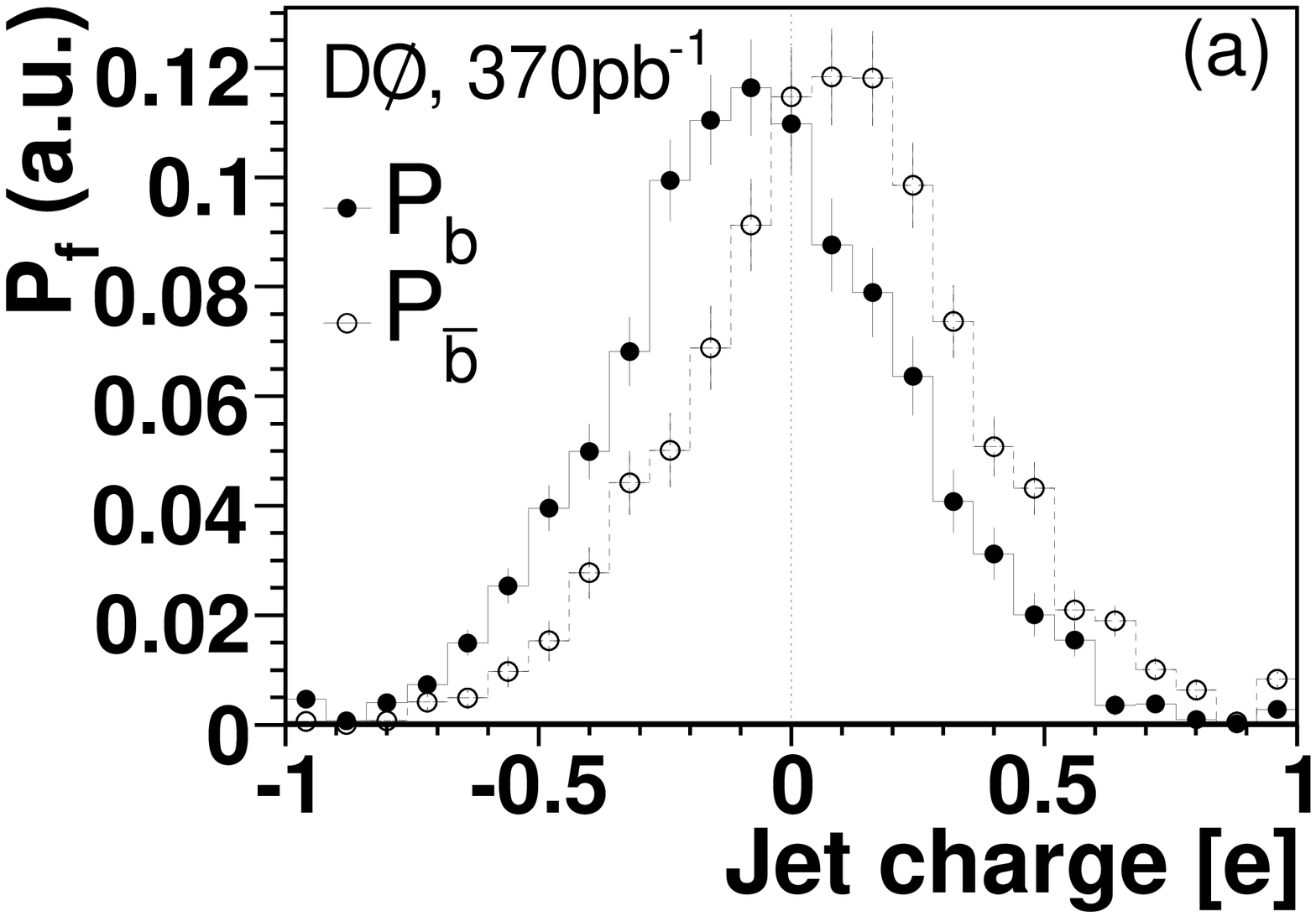}
  \includegraphics[width=0.32\linewidth]{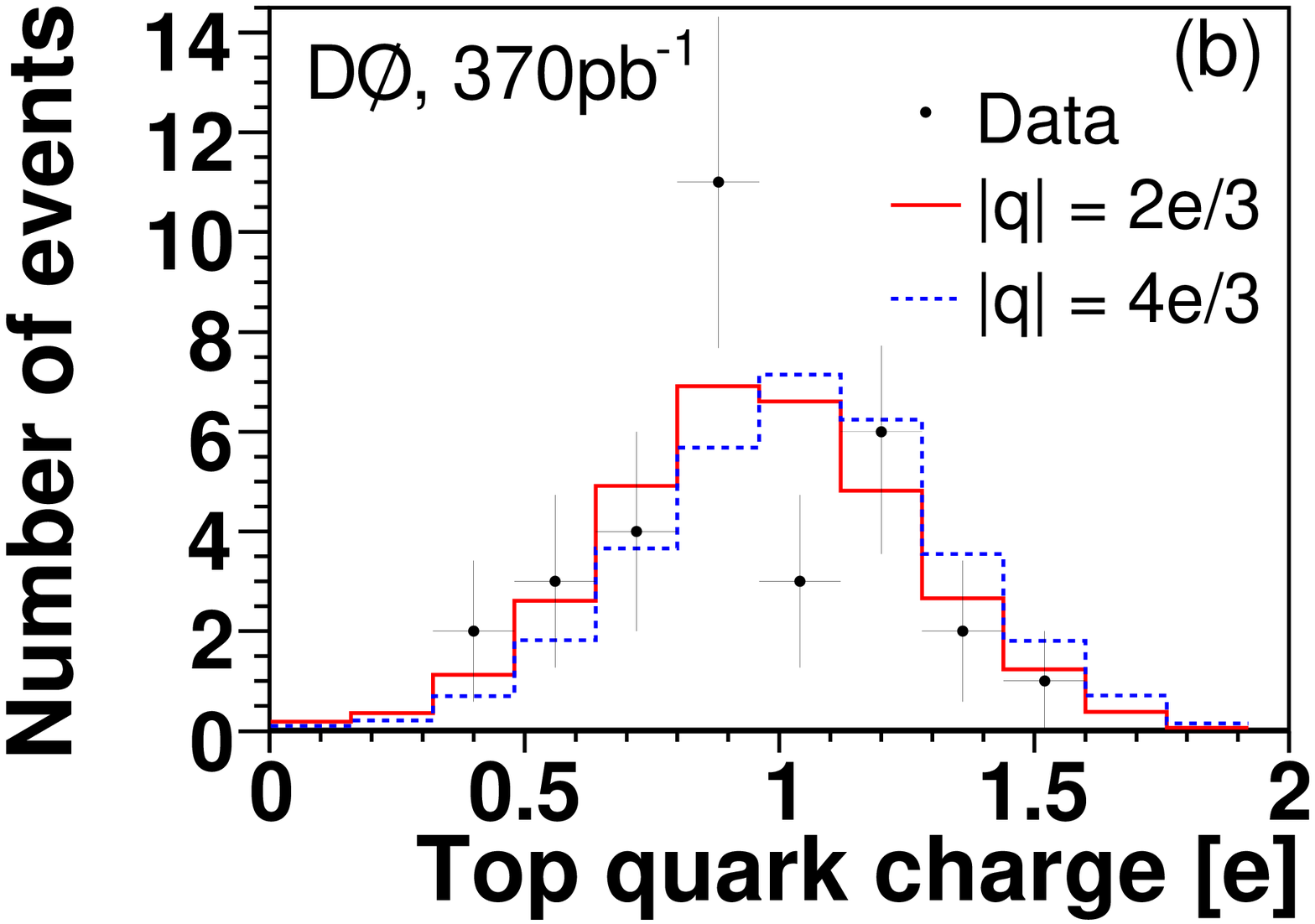}
  \includegraphics[width=0.32\linewidth]{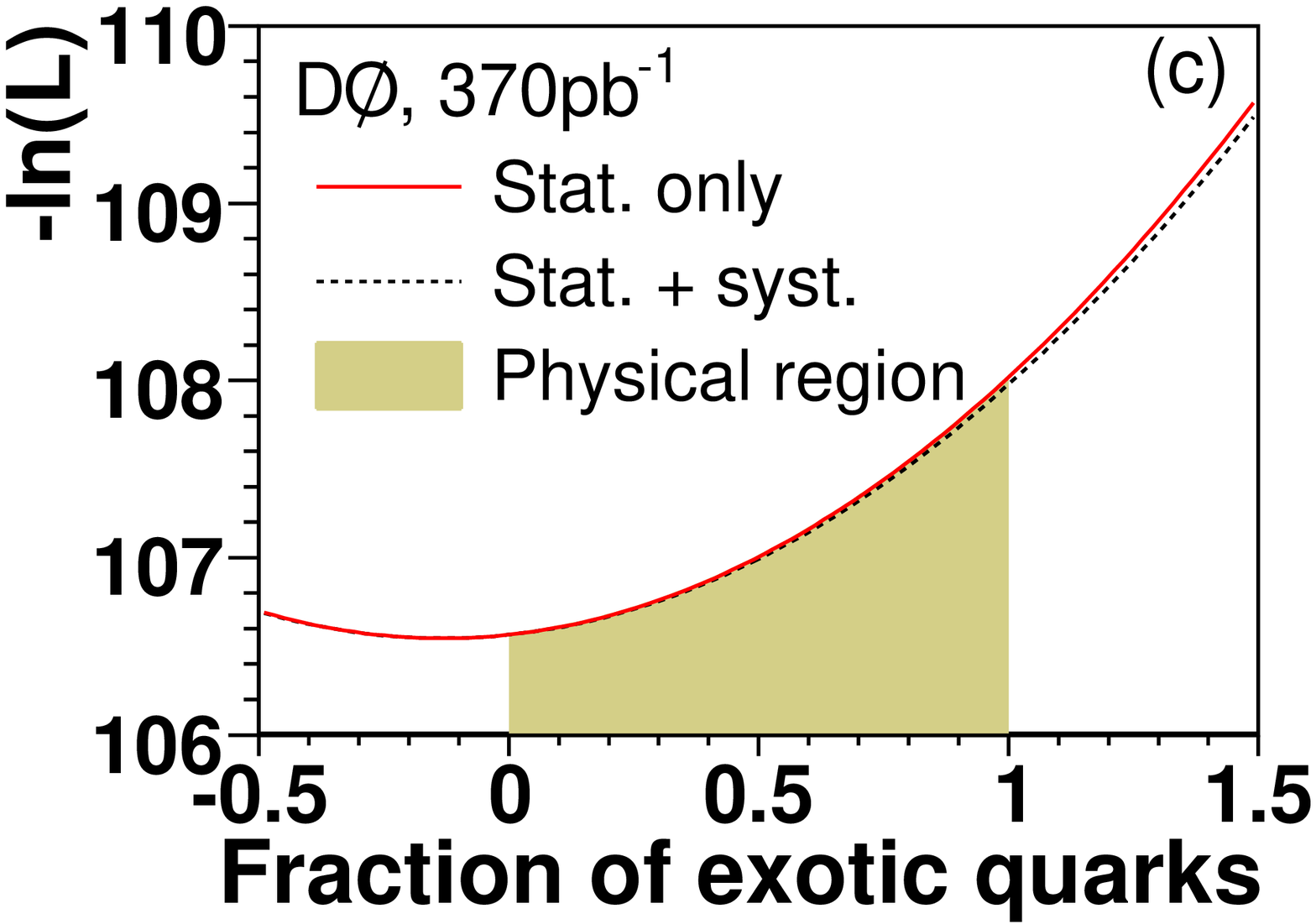}
  \caption{Expected jet charge distribution for $b$ and anti-$b$
    quarks (left). Measured absolute top charge compared to SM and exotic
    models (middle). Likelihood as function of the fraction of exotic
  top quarks (right).}
  \label{fig:charge}
\end{figure}
\subsection{Top Charge Results}
To determine the top charge an
assignment of $b$-jets to the leptonic or hadronic event side is
necessary. This analysis uses the quality of a fit to the $t\bar t$
hypothesis, which uses the $W$ and top masses as constraints, to
select the best possible assignment.
The jet charge for the $b$ jets for the leptonic (hadronic) side,
$q_{b_l}$ ($q_{b_h}$) is then combined with the charge of the measured
lepton $q_l$ to define two top charge values per event:
$Q_\mathrm{lep}=\left|q_l+q_{b_\mathrm{l}}\right|$ and $Q_\mathrm{had}=\left|-q_l+q_{b_\mathrm{h}}\right|$.
The distribution of the measured top charges is compared to MC
templates for the SM and the exotic case, where the exotic case has
been obtained by permuting the jet charge, see \fig{fig:charge}
(middle).

An unbinned likelihood ratio accounting also for remaining background
yields a $p$-value for the exotic case of $7.8\%$ and a Bayes factor
of 4.3.

\section{Search for Top Pair Resonances}
Due to the fast decay of the top quark, no  resonant production of top
pairs is expected within the SM. However, unknown heavy resonances decaying to top
pairs may add a resonant part to the SM production mechanism. 
Resonant production is possible for massive $Z$-like bosons in
extended gauge theories \cite{Leike:1998wr}, Kaluza-Klein states of the gluon
or $Z$ boson~\cite{Lillie:2007yh,Rizzo:1999en}, axigluons
\cite{Sehgal:1987wi}, Topcolor \cite{topc1}, and other theories beyond the SM.  
Independent of the exact model, such resonant production could
be visible in the reconstructed $t\bar t$ invariant mass. 

\subsection{Data Selection, Signal and Background Description}
D0 investigated the invariant mass distribution of top pairs in 
$2.1\ifb$ of $\ell+$jets events\cite{Abazov:2008ny,D0Note5600conf}. The event selection 
and background description follow closely the description in
section \ref{sect:stop.selection}, with the exception that also events
with only three jets are considered.
Signal simulation is created for various resonance masses between $350$
and $1000\GeV$. The width of the resonances was chosen to be $1.2\%$
of their mass, which is much smaller than the detector resolution.

\subsection{Top Pair Invariant Mass}
The top pair invariant mass, $M_{t\bar t}$, is reconstructed directly from the
reconstructed physics objects. A constraint fit as described in
\ref{sect:stop.selection} is not applied. Instead the momentum of the
neutrino is reconstructed from the transverse missing energy, $\met$,
which is identified with the transverse momentum of the neutrino and by solving
$M_W^2=(p_\ell+p_\nu)^2$ for the $z$-component of the neutrino
momentum. $p_\ell$ and $p_\nu$ are the four-momenta of the lepton and
the neutrino, respectively.

\begin{figure}[b]
  \centering
\includegraphics[width=0.46\linewidth]{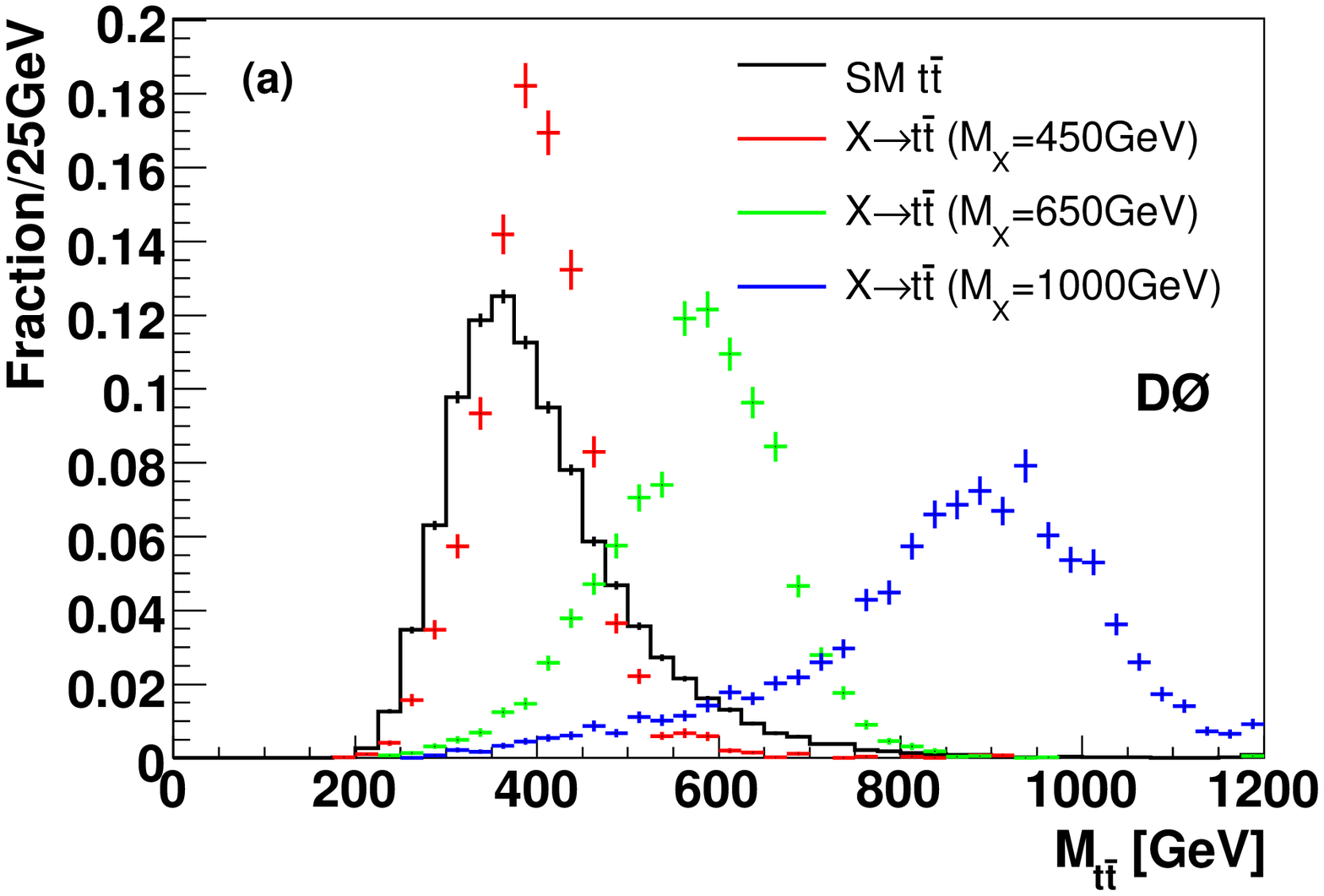}
\includegraphics[width=0.46\linewidth]{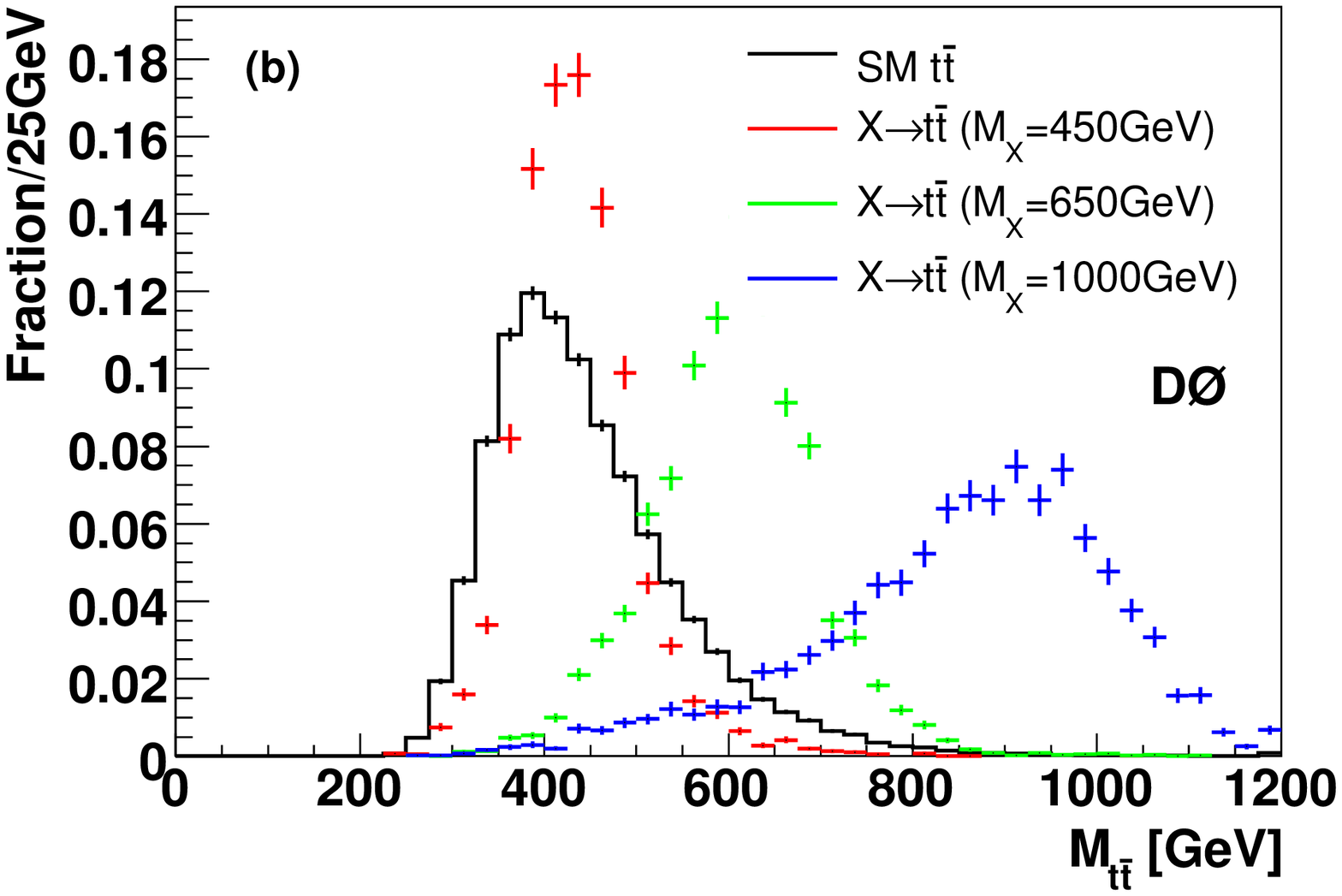}
  \caption{Shape comparison of the expected $t\bar t$ invariant mass distributions for 
SM top quark pair production (histogram)
and 
resonant  production from narrow-width resonances of mass 
$M_X=450$, $650$, and $1000\GeV$, 
for (a) $3$ jets events
and (b)~$\ge 4$~jets events.}
  \label{fig:mtt.mttshapes}
\end{figure}

The $t\bar t$ invariant mass can then be computed without any
assumptions about a jet-parton assignments that is needed in
constraint fits. Compared to the constraint fit reconstruction applied
in earlier analysis this gives  better performance at high
resonance masses and in addition allows the inclusion of $\ell+$3 jets events.
The expected signal shapes for various resonance
masses are compared to the SM top pair distribution in
\fig{fig:mtt.mttshapes}.
The expected distribution of SM processes and the measured data is
shown in \fig{fig:mtt.mtt}.  For comparison a $Z'$ resonance with a mass
of $650\GeV$ is shown at the cross-section expected in the topcolor
assisted technicolor model used for reference.

\begin{figure}[t]
  \centering
\includegraphics[width=0.46\linewidth]{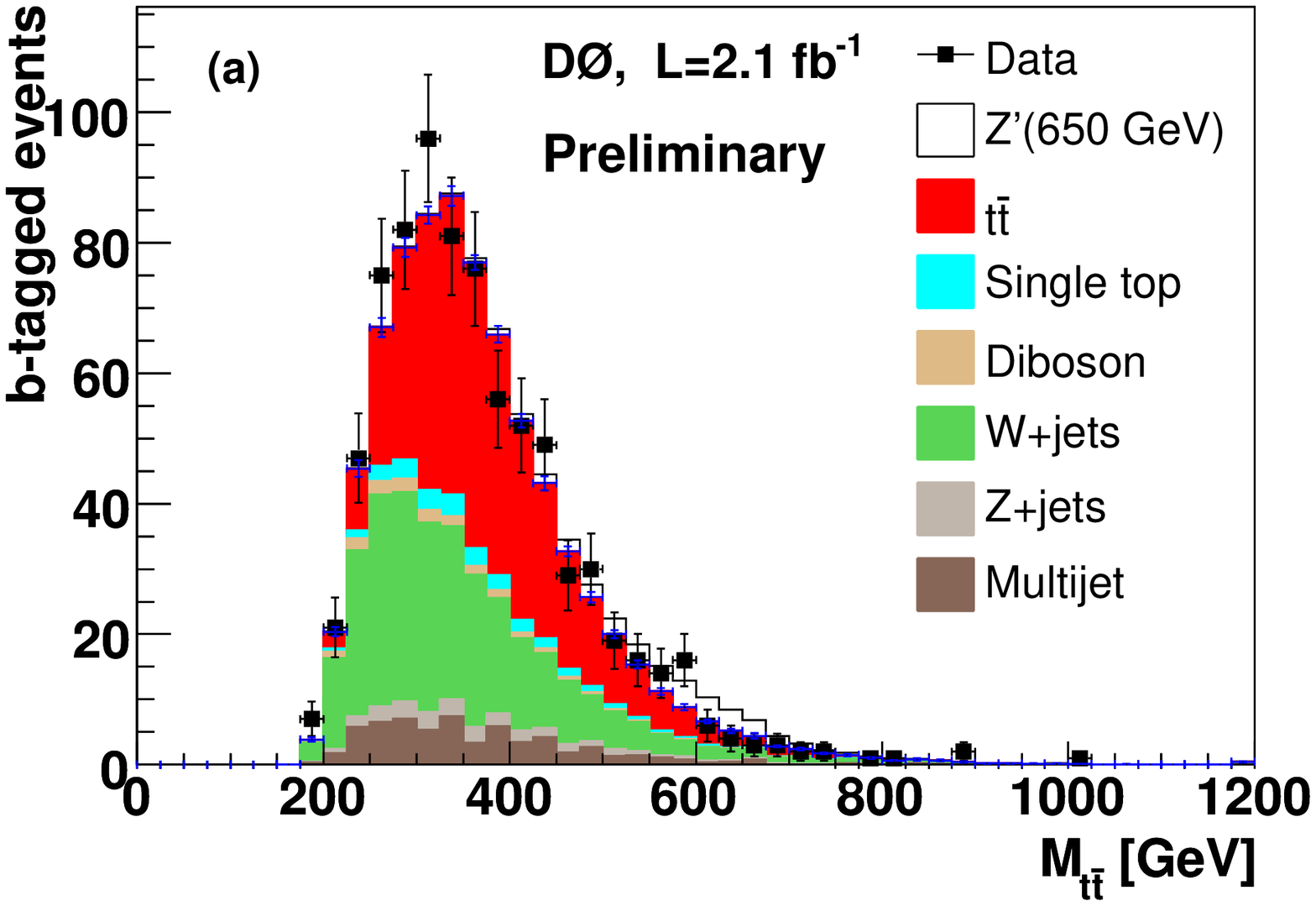}
\includegraphics[width=0.46\linewidth]{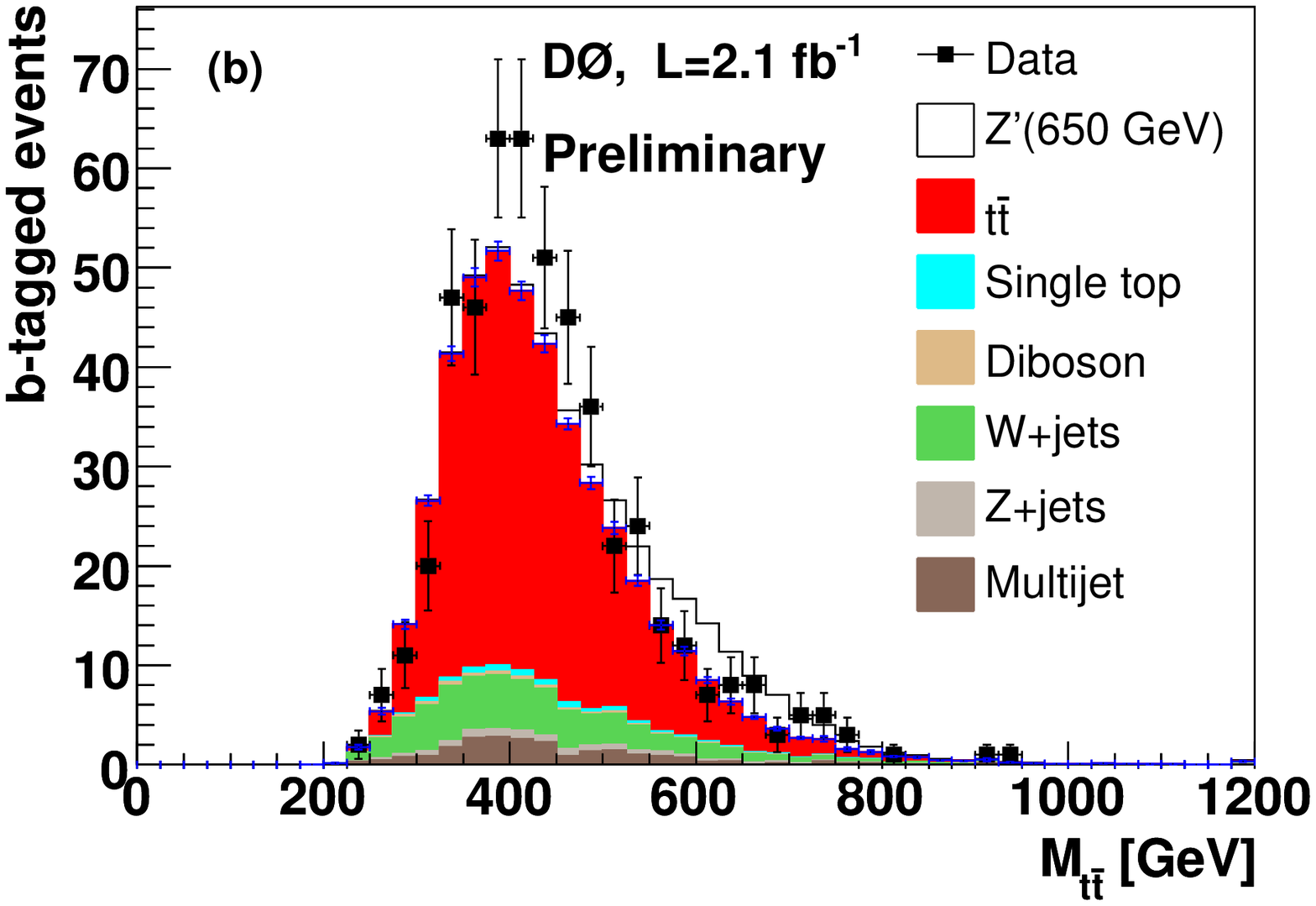}
  \caption{Expected and observed $t\bar t$ invariant mass
  distribution for the combined (a) $\ell+3$~jets and (b) $\ell+4$ or more jets
  channels, with at least one identified $b$ jet.
Superimposed as white area is the expected signal for a
Topcolor-assisted technicolor $Z'$ boson with $M_{Z'}=650\GeV$.}
  \label{fig:mtt.mtt}
\end{figure}

\subsection{Limit Calculation and Systematics}
Cross-section for resonant production are evaluated using the Bayesian
technique described in \ref{sect:stop.limit}. Central values
correspond to the maximum of the posterior probability density, limits
are set at the point where the integral of the posterior probability
density from zero reaches $95\%$ of its total. Expected limits are
obtained by applying the procedure when assuming that the observed
result corresponded to the SM expectation. 

\begin{figure}[b]
  \centering
\includegraphics[width=0.46\linewidth]{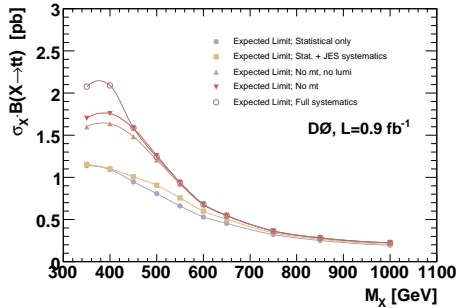}
  \caption{Expected limits (from bottom to top) without systematics, including only JES
systematics, excluding selection efficiencies, $m_t$ and luminosity,
 all  except $m_t$ and complete systematics.
}
  \label{fig:mtt.syst}
\end{figure}
These expected limits were used to optimise major analysis cut and the
$b$-tag working point. In \fig{fig:mtt.syst} the expected limits are
used to visualise the effect of the various systematics by including one
after another. The lowest curve corresponds to a purely statistical
limit. Adding the jet energy uncertainty shows that this uncertainty
mainly contributes at medium resonance masses. The various object
identification efficiencies and the luminosity are added. They
essentially scale like the background shape. Finally the effect of
the top mass is shown and it is most important at low resonance masses.

\subsection{Results}

\begin{figure}[t]
  \centering
  \includegraphics[width=.49\textwidth]{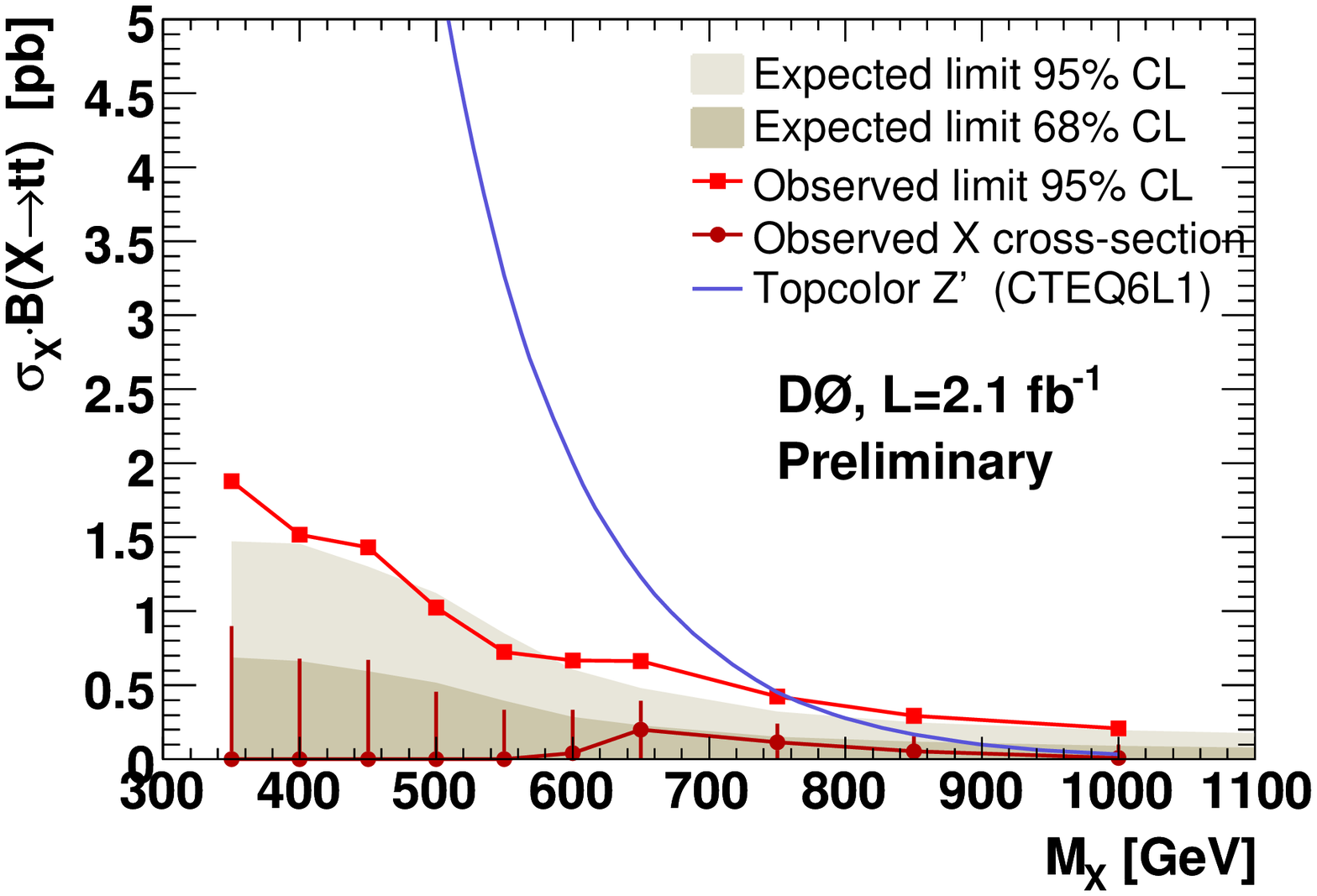}
\includegraphics[width=.49\textwidth]{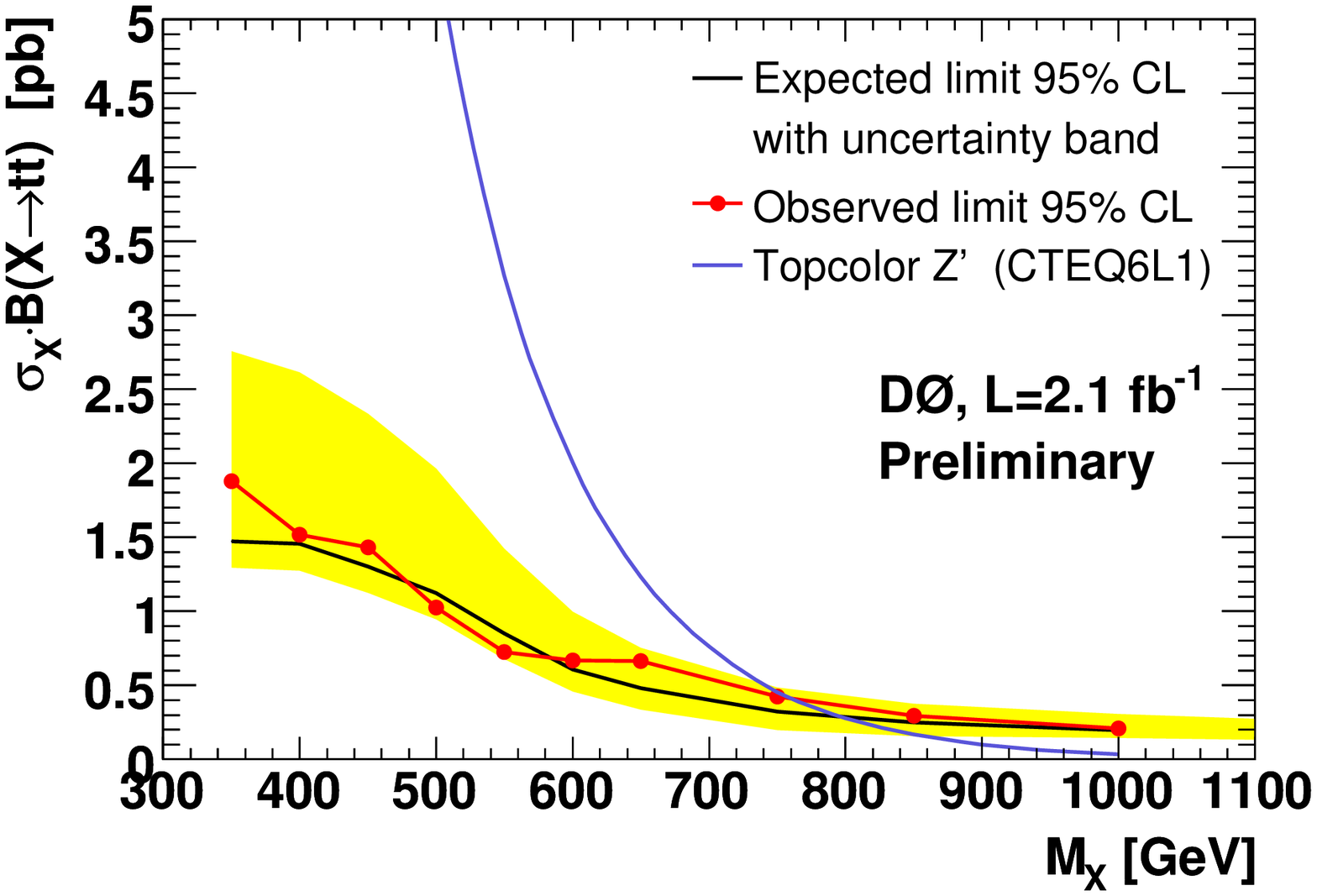}%
  \caption{Left: Expected  upper limits on
  $\sigma_X\cdot B(X\rightarrow t\bar t)$ (shaded) vs.\ 
the assumed resonance mass compared to the observed
cross-section and exclusion limits at
95\%CL.
Right: Expected limit with 1$\sigma$ uncertainty band compared to
observed result.
Both figures show the prediction of the topcolor assisted technicolor
model used to derive benchmark mass limits. 
  }
  \label{fig:mtt.result} 
\end{figure}

The observed cross-sections are close to zero for all considered
resonance masses, as shown \fig{fig:mtt.result} (left). The largest
deviation (around $700\GeV$) is about one standard deviation. Thus limits
are set on the $\sigma_X\cdot B(X\rightarrow t\bar t)$ as function of
the assumed resonance mass, $M_X$. The excluded values range from
less than about $2\pb$ for low mass resonances to less then $0.21\pb$ for the highest
considered resonance mass of $1\TeV$. 
The benchmark topcolor assisted technicolor model can be excluded for
resonance masses of $M_{Z'}<760\GeV$ at $95\%$CL.
\section{Summary}
D0 has searched for signs of physics beyond the standard in top pair
production signatures in various aspects of top quark production,
properties and decays. In datasets of up to $2.1\ifb$ no deviation
from the standard model of elementary particle physics was found.

\bibliographystyle{varenna}
\bibliography{note,local}
\end{document}
\endinput